\documentclass[10pt,prl,aps,twocolumn,showpacs,superscriptaddress,floatfix]{revtex4}
\usepackage{graphicx}
\usepackage{bm}
\usepackage{amssymb}
\usepackage{color,ulem}
\definecolor{bl}{gray}{0.7}

\begin{document}

\title{Does the side jump effect exist?}
\author{O.P. Sushkov}
\affiliation{School of Physics, University of New South Wales, Sydney 2052, Australia}
\author{A. I. Milstein}
\affiliation{Budker Institute of Nuclear Physics, 630090 Novosibirsk, Russia}
\author{M. Mori}
\affiliation{The Advanced Science Research Center, Japan Atomic Energy Agency, Tokai 319-1195, Japan}
\affiliation{CREST, Japan Science and Technology Agency,
Sanbancho 102-0075, Japan}
\author{S. Maekawa}
\affiliation{The Advanced Science Research Center, Japan Atomic Energy Agency, Tokai 319-1195, Japan}
\affiliation{CREST, Japan Science and Technology Agency,
Sanbancho 102-0075, Japan}

\begin{abstract}
The side-jump effect is a manifestation of the spin orbit interaction in electron scattering from 
an atom/ion/impurity.  The effect has  a broad interest because of its conceptual importance for 
generic spin-orbital physics, in particular the effect is widely discussed in spintronics. 
We reexamine the effect  accounting for the exact nonperturbative electron wave function
inside the atomic core. We find that value of the effect is much smaller than estimates accepted in literature.
The  reduction  factor is $1/Z^2$, where $Z$ is the nucleus charge of the atom/impurity.
This implies that the side-jump effect is practically irrelevant for spintronics, the skew scattering and/or 
the intrinsic mechanism always dominate the anomalous Hall and spin Hall effects.
\end{abstract}

\date{\today}
 

\maketitle
There are two well known manifestations of the relativistic spin-orbit interaction in atoms; 
(i) spin orbital splitting of energy levels (fine structure) and 
(ii) right-left asymmetry in scattering (skew scattering)~\cite{LL,Kessler}.
The side jump effect is another manifestation of the spin-orbit interaction.
The effective spin orbit interactions in solids,  Luttinger Hamiltonian~\cite{LK}, 
Dresselhaus~\cite{Dr}, Rashba~\cite{Rashba} and Dzyaloshinskii-Moriya~\cite{DM}
interactions have the same origin as atomic fine structure. 
These lead to intrinsic spin-orbital effects irrelevant to a disorder/impurities in solids.
Scattering of polarized electrons from impurities in a solid  has a skew
component completely analogous to that in atomic physics.
This is an extrinsic spin-orbital effect.

Manipulation and detection of spins without magnetic field is one of crucial 
aspects in spintronics and can be done by the spin current \cite{Maekawa}. 
By the spin Hall effect (SHE), the spin current can be generated from the 
charge current in a paramagnetic material without magnetic field and vice 
versa \cite{Dyakonov,Murakami03,Sinova04,Kato04,Wunder05}. 
The SHE is similar to the anomalous Hall effect (AHE) in ferromagnetic 
materials. The spin-orbit interaction is necessary for both SHE and AHE (for review see 
Refs.~\cite{Sinova06,Engel,Sinitsin,Nagaosa}).
The first mechanism for SHE proposed by Dyakonov and  Perel~\cite{Dyakonov}  was the extrinsic 
one due to the skew scattering from  impurities.
Several intrinsic mechanisms have been also proposed~\cite{Murakami03,Sinova04,Kontani07,Onoda}.

The side jump is another extrinsic mechanism for SHE and AHE.
Remarkably, in spite of its generic importance, the effect is mostly unknown outside of spintronics
community.
The idea of the side jump effect was implicitly formulated in the pioneering work by 
Karplus and Luttinger~\cite{Karplus}. 
Their approach has been further developed by Smit~\cite{Smit}.
Among other mechanisms, Smit considered a possibility of the transverse jump of 
the wave packet. However, according to his analysis, this jump is equal to zero due to 
special kinematic  cancellations. The idea of the transverse coordinate jump of the wave packet 
was reintroduced by Berger, who has suggested the term ``side jump''~\cite{Berger}.
The theory of the side jump effect in its modern form was developed by Lyo and Holstein~\cite{Lyo}.
There are numerous discussions of the implications of the side jump effect 
(see Refs.~\cite{Nozieres,Rashba08,Culcer}). 
It is worth noting some similarity between the side jump effect and 
Imbert-Fedorov effect in optics~\cite{Fedorov}.
\begin{figure}[ht]
\includegraphics[width=0.3\textwidth,clip]{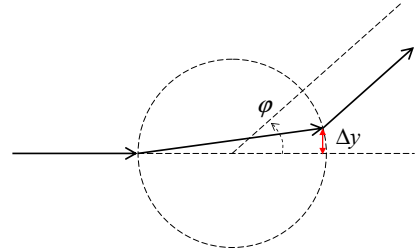}
\caption{Scattering of a polarized electron (polarization perpendicular to the plane)
  from impurity/atom/ion. $\varphi$ is the skew scattering angle (right-left asymmetry)
and $\Delta y$ is the side jump displacement.
}
\label{skewSJ}
\end{figure}
At a cartoon level, the idea of side jump is explained in Fig.~\ref{skewSJ}, which shows
scattering of a polarized electron from an atom. The average deflection angle $\varphi$ describes the
skew scattering (right-left asymmetry) and the average displacement $\Delta y$ is the side jump.
Both the  skew scattering and the side jump contribute to the the multiple scattering deflection.
As one might guess from the cartoon, 
the side jump can be significant compared to the skew
only at a sufficiently high density of scattering centers when the mean free path is short.

For calculations, we adopt the Lyo \& Holstein approach~\cite{Lyo}, in which the side jump  transverse current is 
equal to the expectation value of the correction to the current operator due to the spin-orbit interaction.
All previous calculations of the side jump effect were based on perturbation theory in the scattering 
potential. On the other hand, it is well known that such perturbation theory is not valid for atomic
fine structure and for atomic skew scattering~\cite{LL,Kessler}.
The philosophy previously used for the side jump effect was to
calculate the ratio of the side jump and the skew scattering  within the perturbation theory
and then to assume that nonperturbative effects do not change the ratio.
In the present work, we demonstrate that this assumption is wrong. 
Intra-atomic nonperturbative effects
suppress the ratio of the side jump over skew by orders of magnitude.
The suppression  factor is $1/Z^2$, where $Z$ is the atom/impurity nuclear charge.
This is because the skew scattering scales as $\propto Z^2$ while the side jump is approximately $Z$-independent.
Thus we conclude that  the side jump mechanism is practically irrelevant even at high density of impurities,  though 
the effect itself exists.  

Consider the electron scattering from an impurity/atom.
The Hamiltonian describing the problem is,
\begin{eqnarray}
\label{H}
H&=&H_0+U_{ls},\\
H_0&=&{\bm p}^2/2m +U(r),\nonumber\\
U_{ls}&=& \eta_{ls}\frac{1}{r}\frac{dU}{d r} ({\bm l}\cdot{\bm S}). \nonumber
\end{eqnarray}
Here $U(r)$ is the impurity potential,
${\bm l}=[{\bm r}\times{\bm p}]/\hbar $ and  ${\bm S}$ are orbital momentum 
and the spin of the electron, respectively.
The parameter of spin-orbit interaction is denoted by 
$\eta_{ls}=\hbar^2/(2m^2c^2)$ with the speed of light $c$ and the electron mass $m$.

\begin{figure}[ht]
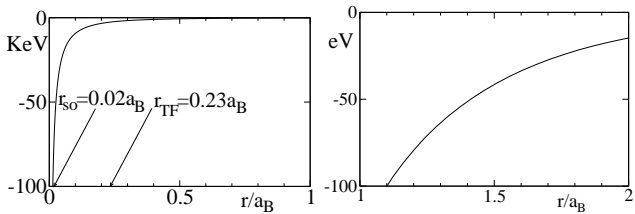

\includegraphics[width=0.23\textwidth,clip]{SJUXe1.eps}
\includegraphics[width=0.23\textwidth,clip]{SJUXe2.eps}
\caption{Potential $U(r)$ in  Xe atom ($Z=54$).
The left panel corresponds to $r < a_B$ with the vertical scale given in KeV.
The right panel corresponds to $a_B< r < 2a_B$ with the vertical scale given in eV.
Arrows in the left panel show the Thomas-Fermi and the spin-orbit scales.
}
\label{impXe}
\end{figure}

To illustrate scales involving in the problem, we present in Fig.~\ref{impXe} the
self-consistent Hartree-Fock potential of Xe  atom~\cite{Dzuba}.
The typical spatial size of an atom/impurity is a few  Bohr radii, $a\sim few \times a_B$,  
$a_B =\hbar^2/(me^2)\approx 0.5 \AA$.
The typical value of $U(a)$ is
about several eV. In Xe atom $U(r=2a_B)\approx -Ry$, where
$Ry=(me^4/\hbar^2)/2=13.6$ eV  is Rydberg.
At $r > a $ the potential quickly decays to zero due to electron screening.
On the other hand, when $r$ is decreasing, $r < a$, the potential grows strongly because of reduced
screening. In Xe  atom $U(r=a_B)\approx -340eV$. 
Below the Thomas-Fermi radius, $r < r_{TF} = 0.885 a_B/Z^{1/3}$~\cite{LL}, the potential
grows even steeper, $U(r) \approx -2 Z Ry (a_B/r)$, because here
nuclear charge is practically unscreened.
This very singular behavior of the potential is evident in Fig.~\ref{impXe}.
In the right panel, $1< r/a_B < 2$, the potential is given in eV, while
in the left panel, $0 < r/a_B < 1$, the potential is given in KeV.

It is well known~\cite{LL} that all spin orbital effects originate from distances $r\sim r_{so} = a_B/Z$.
This scale is indicated 
in the left panel of Fig.\ref{impXe}.
At this distance, the potential is enormous, $U(a_B/Z)= 2Z^2Ry$ (which is 79KeV for Xe), and the wave function 
behaviour is highly
nonperturbative.  This important fact was completely missing in all previous considerations
of the side jump effect.
In previous works the true very singular atomic potential was replaced by a soft nonsingular
pseudopotential $U(r) \to V(r)$ and calculations of both the skew scattering amplitude and 
the side jump
were performed within simple perturbation theory in the pseudopotential $V(r)$.
Any possible dependence on the nuclear charge $Z$ was missing from the very beginning because there
was no such a parameter in the analysis.
On the other hand, it is clear that the nuclear charge is very important, because the spin 
orbit interaction is  negligible in light atoms, $Z\sim 1$, and the interaction is much more 
significant in heavy atoms at $Z \gg 1$.

For low energy electrons, $|\epsilon|\lesssim Ry$, the spin-orbit energy splitting (fine structure) 
scales  as $Z^2$~\cite{LL}. We remind how this important statement is derived.
Let us denote the electron wave function at 
$r\sim a$ by $\psi_0$. A perturbative estimate of the effective spin-orbit 
interaction with a soft pseudopotential $V(r)$ gives,
\begin{eqnarray}
\label{est1}
\langle\psi \left|\eta_{ls}\frac{1}{r}\frac{dV}{d r} ({\bm l}\cdot{\bm S})
\right|\psi\rangle &\sim& \eta_{ls}\frac{Ry}{a_B^2}
\langle\psi_0 |({\bm l}\cdot{\bm S}) |\psi_0\rangle\nonumber\\
&\sim& \alpha^2Ry \ \langle\psi_0 |({\bm l}\cdot{\bm S}) |\psi_0\rangle \ ,
\end{eqnarray}
where $\alpha=e^2/(\hbar c)=1/137$ is the fine structure
constant. Naturally, according to this estimate, the effective spin-orbit 
constant, $\lambda \sim \alpha^2Ry \sim 10^{-3}eV$, is very small and Z-independent.
The true $Z^2$ scaling is a nonperturbative effect. 
Because of the very strong potential, the electron wave function at 
$r \sim r_{so} \sim a_B/Z$ is strongly enhanced compared to its exterior value~\cite{LL} such as, 
\begin{equation}
\label{psi1}
\psi(r) \sim \frac{\psi_0}{Z^{1/4}(r/a_B)^{3/4}} \ .
\end{equation}
In particular $\psi(r_{so})\sim \sqrt{Z}\psi_0$.
Calculation of the matrix element of $U_{ls}$ with the wave function
(\ref{psi1}) shows that the matrix element comes from the distances 
$r \sim r_{so}$ and direct integration gives~\cite{LL}, 
\begin{equation}
\label{estZ}
\langle\psi |U_{ls} |\psi\rangle
\sim (Z\alpha)^2Ry \ \langle\psi_0 |({\bm l}\cdot{\bm S}) |\psi_0\rangle \ .
\end{equation}
Hence the effective spin-orbit constant is $\lambda \sim (Z\alpha)^2 Ry$. 
We disregard the $l$-dependence of $\lambda$ ($l$ is the angular momentum of the electron).  
The $l$-dependence can be easily included~\cite{Haken}, but this is not important for our purposes.
It is worth noting that due to the band structure the electron dispersion at low momenta is different 
from simple parabolic one (\ref{H}). Again, this is not important for us since at small distances
(large momenta) relevant to the problem the dispersion is always quadratic.

The skew scattering is proportional
to the scattering phase shift between the partial waves with total angular momenta
$j=l+1/2$ and $j=l-1/2$~\cite{LL,Kessler}. 
Practically for electrons with $\epsilon \lesssim Ry$ only $l=0,1$ are important
and the scattering cross section reads (assuming $\delta_s \gg \delta_p$)
\begin{eqnarray}
\label{skew1}
\frac{d\sigma}{d\Omega}=\frac{d\sigma_0}{d\Omega}\left\{1+4(\delta_{p_{3/2}}-\delta_{p_{1/2}})
({\bm S}\cdot[{\bm n}\times {\bm n}'])\right\} \ ,
\end{eqnarray}
where $\frac{d\sigma_0}{d\Omega} =|f|^2$, $f \sim a$ is the scattering amplitude
without account of the spin-orbit interaction. Unit vectors
${\bm n}$ and ${\bm n}'$ are directed along initial and final electron momenta 
respectively.
The phase shift can be calculated~\cite{LL} using  Eq. (\ref{estZ})
\begin{equation}
\label{ph1}
\delta_{p_{3/2}}-\delta_{p_{1/2}} \sim -\frac{ m a^2}{\hbar^2}\langle\psi |U_{ls} |\psi\rangle
\sim -(Z\alpha)^2 \ .
\end{equation} 
Therefore the skew cross section is
\begin{eqnarray}
\label{skew2}
\frac{d\sigma}{d\Omega}=\frac{d\sigma_0}{d\Omega}\left\{1-\gamma(Z\alpha)^2
({\bm S}\cdot[{\bm n}\times {\bm n}'])\right\} \ ,
\end{eqnarray}
where $\gamma\sim 1$ is a constant. Eqs. (\ref{ph1}) and (\ref{skew2}) are valid for electron
energy above the centrifugal barrier.  The barrier is always low as 1.8 eV in neutral Xe and zero in ions.
Below the barrier, there is an additional centrifugal suppression,
$\delta_{p_{3/2}}-\delta_{p_{1/2}}\to (Z\alpha)^2(ka)^3$, where
$k \ll 1/a$ is the wave vector of the electron.
Eqs. (\ref{ph1}) and (\ref{skew2}) agree well with direct measurements for electron scattering
from  Xe~\cite{Xedata}.

Thus, similar to the fine structure splitting, the skew scattering scales as $\propto Z^2$.
Therefore, it is important to know if the side jump has the same enhancement.
The operator of the side jump velocity is 
\begin{eqnarray}
\label{comm1}
\delta {\hat {\bm v}}^{(sj)}&=&-\frac{i}{\hbar}[{\bm r},H]-
\frac{{\bm p}}{m}=
\eta_{ls}[{\bm S}\times{\bm \nabla}U] \ .
\end{eqnarray}
This operator is proportional to ${\bm \nabla} U =({\bm r}/r)(dU/dr)$.
So, it has an additional power of the radius $r$ compared
to $U_{ls}$ given by Eq. (\ref{H}).
Therefore, the straightforward {\it upper} estimate of the matrix element of Eq.~(\ref{comm1})
with the wave function (\ref{psi1}) gives only the first power of $Z$, i.e., $\langle \delta {\bm v}\rangle  \propto Z$.
Even this small result is an overestimate,
due to exact equations of motion $\langle \delta {\bm v}\rangle\propto Z^0 \sim 1$.
To prove this we represent ${\bm \nabla} U$  as
\begin{eqnarray}
\label{nU}
{\bm \nabla}U=\frac{i}{\hbar}[{\bm p},H_0]\ .
\end{eqnarray}
Hence to find the expectation value of $\langle \delta {\bm v}\rangle$ from Eq.~(\ref{comm1})
we need to calculate the following matrix element in the limit $\mu \to +0$
\begin{eqnarray}
\label{expmu}
\frac{i}{\hbar}\langle\psi_{{\bm k}+}e^{-\mu r}[{\bm p},H_0]
e^{-\mu r}\psi_{{\bm k}+}\rangle \ .
\end{eqnarray}
Here  $\psi_{{\bm k}+}$ is the scattering state asymptotically, $kr \gg 1$, consisting of the incident plane 
wave and the  diverging spherical wave,
\begin{equation}
\label{psi+}
\psi_{{\bm k}+} \to e^{i{\bm k}\cdot{\bm r}}+\frac{f}{r}e^{ikr},
\end{equation}
where $f$ is the scattering amplitude.
The parameter $\mu \to +0$ is introduced to regularize the matrix element~(\ref{expmu}). 
This regularization is necessary for the wave-packet scattering problem to converge the integrals at $r\to \infty$.
The matrix element (\ref{expmu}) is calculated by commuting $H_0$ with
the regularization factor, 
\begin{eqnarray}
\label{mh}
[H_0,e^{-\mu r}]&=&[p^2/(2m),e^{-\mu r}]\nonumber\\
&=&\frac{i\mu\hbar}{2m}\left({\bm p}\cdot\frac{\bm r}{r}e^{-\mu r}
+e^{-\mu r}\frac{\bm r}{r}\cdot{\bm p}\right).
\end{eqnarray}
Hence, Eq.~(\ref{expmu}) is reduced to
\begin{equation}
\label{mh1}
-\frac{\mu}{m}\mbox{Re}
\langle\psi_{{\bm k}+}\left|\left[({\bm p}\cdot{\bm r})\frac{1}{r}e^{-\mu r}
+e^{-\mu r}\frac{1}{r}({\bm r}\cdot{\bm p})\right]{\bm p}e^{-\mu r}
\right|\psi_{{\bm k}+}\rangle \ .
\end{equation}
Here ``Re'' stands for the real part.
Importantly the expression (\ref{mh1}) is proportional to the
infinitesimally small $\mu$.
Hence, the matrix element in Eq.~(\ref{mh1}) must be calculated only up
to the order $\mu^{-1}$, all the higher orders, $\mu^0$, $\mu^1$, ... will
give zero contributions in the limit $\mu \to 0$.
The terms proportional to $1/\mu$ can appear only from large distances,
$r\sim 1/\mu$. 
Hence, we can use the asymptotic form of the wave function (\ref{psi+})
to calculate the matrix element in Eq.~(\ref{mh1}).
Note that the wave function (\ref{psi+}) is diverging at $r \to 0$.
The divergence is a byproduct of the asymptotic form and therefore
integrals of $r$ in (\ref{mh1}) have a lower cutoff of the order of atomic size.
All the terms which are sensitive to the value of the cutoff disappear
in the limit $\mu \to 0$.

Here, one can notice two interesting points.
The side jump current, which is equal to the matrix element of
Eq.~(\ref{comm1}), flows at very small distances from the nucleus, $r \sim r_{so} \sim a_B/Z$.
Nevertheless the exact equation of motion (\ref{nU}) allows us to translate
calculation of the integrated side jump current to the large distances, 
$r\sim 1/\mu \to \infty$.
The second point concerns the infrared regularization $\mu$.
The first order  perturbation theory calculation ($U(r)$ is the perturbation), which was performed in Ref.~\cite{Lyo},
also uses an infrared regularization by doing the substitution
$1/(\epsilon-\epsilon_{k}) \to 1/(\epsilon-\epsilon_{k}+i\mu) \to  -i \pi \delta(\epsilon-\epsilon_{k})$.
Our regularization method (\ref{mh1}) is exact, and does not refer to the perturbation theory.
We cannot rely on the perturbation theory, since we account all orders in $U(r)$.
 
In the order of $1/\mu$, the both terms in the 
square brackets in Eq.~(\ref{mh1}) give equal contributions and Eq.~(\ref{mh1})
is transformed to
\begin{equation}
\label{cmh1}
\frac{\hbar^2}{m}\left\{4\pi k {\mbox {Im}}f(0,0) \, {\bm n}
-k^2\int d\Omega|f(\theta,\varphi)|^2{\bm n}'\right\} \ .
\end{equation}
We remind that ${\bm n}=(0,0,1)$ is the unit vector along momentum
of the incident electron, $f(\theta,\varphi)$ is the scattering amplitude, 
$d\Omega$ is the scattering solid angle, and 
${\bm n}'=(\sin{\theta}\cos{\varphi},\sin{\theta}\sin{\varphi},\cos{\theta})$
is the unit vector in the scattering direction.
Due to the optical theorem the imaginary part of the forward scattering 
amplitude is related to the total cross section~\cite{LL},
\begin{equation}
\label{opt}
{\mbox {Im}}f(0,0)=\frac{k}{4\pi}\sigma_{0}
=\frac{k}{4\pi}\int d\Omega|f(\theta,\varphi)|^2\ .
\end{equation}
Here the scattering amplitude $f$ and the scattering cross section $\sigma_0$
are calculated without account of the spin-orbit interaction.
We remind that while we treat the interaction potential {\it exactly},
the spin-orbit interaction is considered only in the first order.
Hence Eq.(\ref{cmh1}) is transformed to the transport cross section and we
obtain the following relation
\begin{eqnarray}
\label{Uf}
&&\langle\psi_{{\bm k}+}|{\bm \nabla}U|\psi_{{\bm k}+}\rangle=
{\bm k}\frac{\hbar^2k}{m}\sigma_{tr}=\hbar {\bm k} v \sigma_{tr}\\
&&\sigma_{tr}=\int (1-\cos\theta) |f(\theta,\varphi)|^2d\Omega \ ,\nonumber
\end{eqnarray}
where $v$ is speed of the electron.
This equation represents the force-momentum balance, i.e., the average force 
acting on the atomic nucleus (the left hand side of Eq.~(\ref{Uf}))
is equal to the momentum transfer from the incident electron beam
(the right hand side of Eq.~(\ref{Uf})). 
Because of this reasoning, one can skip  the technical
derivation of Eqs.(\ref{nU})-(\ref{opt}) and go straight to Eq.~(\ref{Uf}).

Taking the matrix element of Eq.~(\ref{comm1}) over the state $\psi_{{\bm k}+}$
and using the exact relation (\ref{Uf}) we find
\begin{eqnarray}
\label{v33}
\langle \delta{\bm v}^{(sj)}\rangle =\eta_{ls}[{\bm S}\times{\bm k}] v\sigma_{tr} \ .
\end{eqnarray}
This is similar to the result in Ref.~\cite{Lyo}. However (\ref{v33}) has been derived exactly
for arbitrary strong potential of a heavy atom. The most important issue is that the side jump 
(\ref{v33}) does not scale with nuclear charge $Z$, while the skew scattering (\ref{skew2})
scales as $Z^2$ (see comment~\cite{note2}).

The spin Hall conductivity $\sigma_{xy}$ is defined by equation $J_{sy}=\sigma_{xy}E_x$, 
where $J_{sy}$ is the $y$-component of spin current and 
$E_x$ is the $x$-component of electric field. 
The ratio of the side jump (SJ) contribution and the skew 
scattering (SS) contribution to the conductivity was calculated in Ref.~\cite{maekawa06} within 
the perturbation theory. 
Using solution of the kinetic equation from Ref.~\cite{maekawa06}, and using
exact Eqs. (\ref{skew2}) and (\ref{v33}) instead of the corresponding perturbative
Eqs. in Ref.~\cite{maekawa06}, we find
\begin{equation}
\label{SJSS}
\frac{\sigma_{xy}^{SJ}}{\sigma_{xy}^{SS}} \sim \frac{1}{Z^2}\frac{\hbar}{\tau_{tr} Ry } \ .
\end{equation}
Thus, even in the most dirty metal where $ \frac{\hbar}{\tau_{tr} Ry }\sim 1$, the side jump contribution
to the spin Hall conductivity is suppressed compared to that of the skew scattering by the factor $1/Z^2$.

{\it In conclusion.}
The skew scattering from an impurity/atom (the right-left asymmetry)
is a spin-orbital effect enhanced by the nuclear charge  of the impurity $\propto Z^2$.
The side jump effect is another manifestation of the spin-orbit interaction in
electron scattering from  impurities/atoms.
We have demonstrated that due to the complex intra-atomic structure of the 
electron wave function the side jump effect is not enhanced by $Z$
contrary to the view accepted previously. 
This implies that relative to the skew the side jump effect is 
smaller by the factor $1/Z^2$  compared to all previous estimates. For typical semiconductors,
$Z \sim 30 \ - \ 50$, the suppression factor is about $\sim 10^3$. 
This makes the side jump effect irrelevant compared to the skew scattering even in most dirty materials.

We acknowledge communications with A. I. Frank, and we a thankful to V. A. Dzuba for providing
data for atomic Hartree-Fock potential.
 This work was partly supported by Grant-in-Aid for Scientific Research from MEXT (Grant No. 24540387, No. 24360036, No. 23340093) and by the inter-university cooperative research program of IMR, Tohoku University. 
O.P.S thanks the Japan Society for Promotion of Science and the Advanced Science Research Center, 
Japan Atomic Energy Agency for financial support and kind hospitality. 
The work of A.I.M. was supported  by the Ministry of Education and Science of the  Russian Federation.
A.I.M and M.M. thank the Godfrey Bequest and School of Physics at University of New South Wales for financial support and kind hospitality.


\begin{thebibliography}{99}
\bibitem{note2} The cross section $\sigma_0 \sim \sigma_{tr}\sim \pi a^2 $ only weakly depends on $Z$.
This dpendence is common for both the side jump (\ref{v33}) and for the skew (\ref{skew2}).
\end{thebibliography}
\end{document}